\documentclass[]{hhr_arxiv}

\title{THE IMPLEMENTABILITY OF LIBERALISM}

\newcommand{\myauthorA}{\textsc{H\'{e}ctor Hermida-Rivera}}%
\newcommand{\myemailA}{\href{mailto:hermida.rivera.hector@gtk.bme.hu}{\texttt{hermida.rivera.hector@gtk.bme.hu}}}%
\newcommand{\myaffiliationA}{\linespread{1.2}\selectfont\emph{Quantitative Social and Management Sciences Research Centre, Faculty of Economic and Social Sciences, Budapest University of Technology and Economics, Muegyetem rkp. 3, Budapest H‑1111, Hungary.}
}%

\begin{document}

\titlesolo

\begin{abstract}
This note shows that under the unrestricted domain, there exists a \hyperref[l]{choice liberal} and \hyperref[ni]{Nash implementable} social choice rule if and only if there are at least three players and the outcome set is at least twice as large as the player set. A social choice rule is \hyperref[l]{choice liberal} if and only if for every player, there exists at least one pair of outcomes such that if this player strictly prefers one over the other, the one he prefers is socially desirable and the other one is not. A social choice rule is \hyperref[ni]{Nash implementable} if and only if there exists a mechanism such that at every preference profile, the set of Nash equilibrium outcomes coincides with the set of socially desirable ones. The proof constructs an intuitive \hyperref[ni]{Nash implementing} mechanism.
\end{abstract}

\keywords{liberalism, Nash equilibrium, mechanism, implementation.}

\jelcodes{D71, D82}

\wordcount{4,208}


\section{Introduction}\label{sec:int}

\textsc{In a very influential paper}, \textcite[Theorem I, p. 153]{sen_70} proved that \emph{liberalism} \parencite[Condition L, p. 153]{sen_70} is incompatible with the \emph{Pareto principle} \parencite[Condition P, p. 153]{sen_70} and the \emph{unrestricted domain} \parencite[Condition U, p. 153]{sen_70}. According to \emph{\citeauthor{sen_70}'s liberalism}; for every player, there must exist at least one pair of outcomes such that if this player prefers one outcome over the other, society shall as well. This notion of \emph{liberalism} captures the idea that every society's member should have---at least---some freedom of choice irrespective of others' preferences.

In another landmark paper, \textcite{maskin_99}\footnote{Although \citeauthor{maskin_99}'s \citeyearpar{maskin_99} paper was published in 1999, it circulated as an unpublished manuscript since 1977 \parencite[Acknowledgements, p. 37]{maskin_99}.} provided an almost complete characterisation of the set of attainable outcomes as a function of players' preferences when players choose \textcite{nash_50,nash_51} equilibrium strategies. The key object in \citeauthor{maskin_99}'s \citeyearpar{maskin_99} paper is a \emph{social choice rule}: namely, a correspondence specifying the set of socially desirable outcomes for each preference profile of society's members. In turn, a social choice rule is \hyperref[ni]{Nash implementable} if and only if there exists a mechanism such that at every preference profile, the set of Nash equilibrium outcomes coincides with the set of socially desirable ones. \citeauthor{maskin_99}'s (\citeyear{maskin_99}, Theorem 2 \& 3, pp. 30 \& 31) results assert that a condition of \emph{monotonicity}\footnote{A social choice rule is \emph{monotonic} if and only if every socially desirable outcome remains so whenever it does not fall in any player's ranking relative to the other outcomes. See \textcite[p. 28]{maskin_99}.} is necessary for \hyperref[ni]{Nash implementation}, and that \emph{monotonicity} and a condition of \emph{no-veto power}\footnote{A social choice rule satisfies \emph{no-veto power} if and only if any outcome that is top-ranked by at least all but one players is socially desirable. See \textcite[p. 31]{maskin_99}.} are together sufficient for \hyperref[ni]{Nash implementation} when there are at least three players.

This note explores whether society can be liberal by verifying if a new form of liberalism, called \hyperref[l]{choice liberalism}, is compatible with \hyperref[ni]{Nash implementability} under the unrestricted domain. Therefore, this note introduces a novel form of \emph{liberalism}, loosely inspired by \citeauthor{sen_70}'s \citeyearpar{sen_70}; and substitutes the \emph{Pareto principle} for \hyperref[ni]{Nash implementability}. While \hyperref[ni]{Nash implementability} was originally defined on social choice rules, \textcite{sen_70} originally defined liberalism as a condition on social decision functions. Clearly, there exist different ways to formalise the idea of liberalism for social choice rules. On the one hand, it is possible to define a social decision function that satisfies \emph{Sen's liberalism} and pick, for each societal preference, its set of top-ranked outcomes. On the other, it is possible to formalise liberalism directly as a property on social choice rules. 

When doing so, one natural approach is to ensure that for each player, there exists at least one pair of outcomes such that if this player prefers one over the other, the one he prefers is socially desirable. Such a notion would match \citeauthor{berlin_58}'s (\citeyear{berlin_58}, p. 160) idea of \emph{positive liberty} (i.e., \emph{freedom to get something done}). Another natural possibility is to ensure that for each player, there exists at least one pair of outcomes such that if this player prefers one over the other, the one he disprefers is not socially desirable. Such a notion would match \citeauthor{berlin_58}'s (\citeyear{berlin_58}, p. 156) idea of \emph{negative liberty} (i.e., \emph{freedom from getting something done}).

This note introduces a novel notion of liberalism for social choice rules by which for every player, there exists at least one pair of outcomes such that this player is \emph{decisive} over that pair: namely, if this player strictly prefers one outcome over the other, the one he prefers is socially desirable and the other one is not.\footnote{Under the unrestricted domain condition, this paper's notion of \hyperref[l]{choice liberalism} is incompatible with no-veto power. To see so, consider a preference profile in which all but one players have a unique top-ranked outcome. Then, by no-veto power, such an outcome must be socially desirable. Now, suppose that said outcome is ranked strictly last by the remaining player, and suppose that such a player is decisive over a pair containing that outcome. Then, by \hyperref[l]{choice liberalism}, this outcome cannot be socially desirable. Finally, by the unrestricted domain condition, this preference profile belongs to the domain of the social choice rule. Thus, it is not possible to rely on \citeauthor{maskin_99}'s (\citeyear{maskin_99}, Theorem 3, p. 31) result to prove this note's result.} Therefore, this paper's notion of liberalism, called \hyperref[l]{choice liberalism}, captures \citeauthor{berlin_58}'s \citeyearpar{berlin_58} two notions of liberty. Moreover, while \emph{Sen's liberalism} is a property used to define a societal preference, it is silent with respect to which outcomes society shall actually choose. This paper's definition of liberalism, however, says which outcomes \emph{shall} (and \emph{shall not}) be chosen by society. Hence, this paper's notion of liberalism is---in a sense---stronger than \citeauthor{sen_70}'s \citeyearpar{sen_70}. 

This note's result, in \Cref{th.cl}, asserts that under the unrestricted domain, a \hyperref[l]{choice liberal} and \hyperref[ni]{Nash implementable} social choice rule exists if and only if there are at least three players and the outcome set is at least twice as large as the player set. While a completely negative result would imply that society cannot uphold liberal values, a positive result is not enough to assert the opposite. This conclusion stems from the fact that both \citeauthor{sen_70}'s \citeyearpar{sen_70} and this paper's definition of liberalism are simply a minimal consequence---rather than a full characterisation---of liberal values. Thus, this note's result shall be understood as an indication that society might be able to respect liberal values under certain circumstances. Hence, the most important part of this note's result is perhaps the impossibility for the two-player case, which is the leading case in contractual applications. 

The proof \Cref{th.cl} has three different parts. The first part shows that with less outcomes than twice as many players, no \hyperref[l]{choice liberal} and \hyperref[ni]{Nash implementable} rule exists. In this case, there will be at least two players that will be decisive over two pairs of outcomes containing a shared outcome. Then, whenever one player strictly prefers the shared outcome over the other outcome in his pair, and the other player strictly prefers the other outcome in his pair over the shared outcome; \hyperref[l]{choice liberalism} implies that the shared outcome \emph{must} (and \emph{must not}) belong to the image of the social choice rule. Hence, no \hyperref[l]{choice liberal} rule can exist in this case.

The second part shows that with just two players, no \hyperref[l]{choice liberal} and \hyperref[ni]{Nash implementable} rule exists. In this case, any mechanism that \hyperref[ni]{Nash implements} a \hyperref[l]{choice liberal} rule must have a unique player who can attain any outcome in the image of the social choice rule regardless of his opponent's choice (i.e., \emph{a dictator}). Then, it suffices to consider a preference profile in which the dictator's favourite outcome should not be in the image of the rule, and for which his opponent cannot attain a better outcome for himself.

The third part constructs a simple \hyperref[ni]{Nash implementing} mechanism in which all players simultaneously name one player and one outcome from the pair they are decisive over. When at least all but one players name the same player, the outcome named by the commonly chosen player is the final outcome; else, the final outcome is the one named by the player with the lowest index. In any Nash equilibrium, any player whose chosen outcome is the final outcome must be choosing his preferred outcome within the pair he is decisive over. And for every outcome in the image of the rule, there exists a Nash equilibrium in which all players are naming the player who is decisive over some pair containing that outcome. 

The rest of the note is organised as follows: \Cref{sec.env} defines the environment, \Cref{sec.mech} introduces the notion of a mechanism, \Cref{sec.ax} presents the axioms, \Cref{sec.th} states and proves the result, and \Cref{sec.ex} discusses some intuitive examples.

\section{The Environment}\label{sec.env}

The \emph{environment} is a $4$-tuple $(N,\Omega,\mathcal{R},\sigma)$, where $N=\{1,\dots,n\}$ is a \emph{finite player set}, with $n\geqslant2$; $\Omega$ is a \emph{non-empty outcome set};
\begin{gather}
    \mathcal{R}=\bigl\{R\mid R=(R_i)_{i\in N}:R_i\text{ is player $i$'s \emph{weak ordering} over }\Omega\bigr\}
\end{gather}
is the \emph{set of all weak preference profiles},\footnote{A \emph{weak ordering} is a \emph{complete}, \emph{transitive} and \emph{reflexive} binary relation. A binary relation $R_i$ is \emph{transitive} if and only if for all outcomes $\omega,\omega',\omega''\in\Omega$, $\omega R_i\omega'$ and $\omega'R_i\omega''$ imply $\omega R_i\omega''$; it is \emph{complete} if and only if for all outcomes $\omega,\omega'\in\Omega$, $\omega R_i\omega'$ or $\omega'R_i\omega$; and it is \emph{reflexive} if and only if for all outcomes $\omega\in\Omega$, $\omega R_i\omega$. See \textcite[Definition 1.B.1, p. 6]{mascolelwhinstongreen_95}.} where $\omega R_i\omega'$ if and only if player $i$ \emph{weakly prefers} $\omega$ over $\omega'$, $\omega P_i\omega'$ if and only if player $i$ \emph{strictly prefers} $\omega$ over $\omega'$, and $\omega I_i\omega'$ if and only if player $i$ \emph{is indifferent} between $\omega$ and $\omega'$; and $\sigma:\mathcal{R}\to 2^{\Omega}\backslash\{\emptyset\}$ is a \emph{social choice rule} that associates each preference profile with some non-empty desirable outcome set.\footnote{Together, the assumptions that $\text{dom}(\sigma)=\mathcal{R}$ and $\sigma(R)\neq\emptyset$ for all preference profiles $R\in\mathcal{R}$ can be seen as an analogue of the \emph{unrestricted domain} condition of \textcite[Condition U, p. 153]{sen_70} for social choice rules rather than social decision functions.}

Let $\Sigma=\{\sigma\mid\sigma\text{ is a \emph{social choice rule}}\}$.

\section{Mechanisms}\label{sec.mech}

An \emph{$n$-player mechanism} (or \emph{$n$-player game form}) is a $2$-tuple $\gamma=(S,\eta)$, where $S=\prod_{i\in N}S_i$ is a \emph{strategy profile space} in which for all players $i\in N$, the \emph{strategy space} $S_i$ is non-empty; and $\eta:S\to\Omega$ is an \emph{outcome function} that links each strategy profile with some outcome.

Let $\Gamma=\{\gamma\mid\gamma\text{ is a \emph{mechanism}}\}$.

Consider any mechanism $\gamma\in\Gamma$. Then, given any strategy profile $s\in S$, let $s_{-i}=(s_j)_{j\in N\backslash\{i\}}\in S_{-i}=\prod_{j\in N\backslash\{i\}}S_j$ be the \emph{strategies chosen by all players other than $i$}, and let $\eta(S_i,s_{-i})=\{\omega\in\Omega\mid(\exists s_i\in S_i)(\eta(s_i,s_{-i})=\omega)\}$ be \emph{player $i$'s set of attainable outcomes when the other players choose strategy profile $s_{-i}$}.\footnote{See \textcite[p. 248]{maskinsjostrom_02} for a further discussion on these concepts.}

A \emph{non-cooperative game} is a pair $(\gamma,R)\in\Gamma\times\mathcal{R}$.

\section{The Axioms}\label{sec.ax}

\Cref{sec.ax} introduces this note's axioms: \hyperref[l]{choice liberalism} and \hyperref[ni]{Nash implementability}. The axiom of \hyperref[l]{choice liberalism} can be seen as a strong adaptation of \emph{liberalism} as defined by \textcite[Condition L, p. 153]{sen_70} but for social choice rules rather than for social decision functions in the sense of \textcite{arrow_50,arrow_51}. Hereafter, a player is \emph{decisive} over a pair of outcomes if and only if for all preference profiles in which this player strictly prefers one outcome over the other, the one he prefers is socially desirable and the other one is not.

\begin{axiom}[Choice liberalism]\label{l}
    A rule $\sigma:\mathcal{R}\to2^\Omega\backslash\{\emptyset\}$ is \emph{choice liberal} if and only if for every player, there exists at least one pair of outcomes over which this player is decisive. Formally, if and only if
\begin{gather}
    (\forall i\in N)(\exists\omega,\omega'\in\Omega)(\forall R\in\mathcal{R})[(\omega P_i\omega')\Rightarrow ((\omega\in\sigma(R))\wedge(\omega'\notin\sigma(R)))]
\end{gather}
\end{axiom}

Let $\Sigma_l=\{\sigma\in\Sigma\mid\sigma\text{ is \hyperref[l]{choice liberal}}\}$.

Intuitively, a \hyperref[l]{choice liberal} rule is one that gives every player some sphere of personal freedom: namely, it lets every player choose whether to engage or not in \emph{at least one} activity irrespective of the preferences of other society's members. For example, if a player wants to read a book, he shall be granted the freedom to do so; and if he does not want to read it, he shall be granted freedom to not do so. See \Cref{sec.ex} for detailed examples.

The axiom of \hyperref[ni]{Nash implementability} is borrowed without change from \textcite[p. 25]{maskin_99}. Given any non-cooperative game $(\gamma,R)\in\Gamma\times \mathcal{R}$, a strategy profile $s\in S$ is a \emph{Nash equilibrium} if and only if no player can unilaterally and profitably deviate. Formally, if and only if
\begin{gather}
    (\forall i\in N)(\forall s_i'\in S_i)(\eta(s)R_i\eta(s_i',s_{-i}))
\end{gather}
Let $H(\gamma,R)=\{s\in S\mid s\text{ is a \emph{Nash equilibrium} of }(\gamma,R)\}$.

\begin{axiom}[Nash Implementability]\label{ni}
    \parencite[p. 25]{maskin_99}. A rule $\sigma:\mathcal{R}\to2^\Omega\backslash\{\emptyset\}$ is \emph{Nash implementable} if and only if there exists some mechanism such that at every preference profile, the set of Nash equilibrium outcomes coincides with the set of socially desirable ones. Formally, if and only if
\begin{gather}\label{eq1a}
    (\exists \gamma\in \Gamma)[(\forall R\in \mathcal{R})(\eta(H(\gamma,R))=\sigma(R))]
\end{gather}
\end{axiom}

Let $\Sigma_n=\{\sigma\in\Sigma\mid\sigma\text{ is \hyperref[ni]{Nash implementable}}\}$.

\section{The Theorem}\label{sec.th}

\begin{theorem}\label{th.cl}
    A \hyperref[l]{choice liberal} and \hyperref[ni]{Nash implementable} social choice rule exists if and only if there are at least three players and the outcome set is at least twice as large as the player set. Formally,
\begin{gather}
    (\Sigma_l\cap\Sigma_n\neq\emptyset)\iff[(n\geqslant 3)\wedge(|\Omega|\geqslant 2n)]
\end{gather}
\end{theorem}

\begin{proof}
    The are two statements to show: one for each direction.

\begin{statement}
    If $n<3$ or $|\Omega|<2n$, then $\Sigma_n\cap\Sigma_l=\emptyset$.
\end{statement}

\begin{claim}
    If $|\Omega|< 2n$, then $\Sigma_l\cap\Sigma_n=\emptyset$.
\end{claim}

    The proof is by contradiction. Consider any \hyperref[l]{choice liberal} rule $\sigma\in\Sigma_l$ and let $|\Omega|<2n$. Since $|\Omega|<2n$, there exist two players $i,j\in N$ and some outcome $\omega^*\in\Omega$ such that player $i$ is decisive over some pair $\{\omega^*,\omega'\}$, and player $j$ is decisive over some pair $\{\omega^*,\omega''\}$. Let the preference profile $R\in\mathcal{R}=\text{dom}(\sigma)$ satisfy $\omega^* P_i\omega'$ and $\omega'' P_j\omega^*$. By the \hyperref[l]{choice liberalism} axiom, it follows that $\omega^* P_i\omega'$ implies $\omega^*\in\sigma(R)$; whereas $\omega'' P_j\omega^*$ implies $\omega^*\notin\sigma(R)$, which is a contradiction. Hence, $\Sigma_l=\emptyset$. Consequently, $\Sigma_l\cap\Sigma_n=\emptyset$.

\begin{claim}
    If $n=2$, then $\Sigma_n\cap\Sigma_l=\emptyset$.
\end{claim}

    The proof has three steps. Let $N=\{1,2\}$, let $|\Omega|\geqslant4$ and let $\Omega=\{\omega_1,\dots,\omega_m\}$. Then, consider some mechanism $\gamma\in\Gamma$ satisfying $\eta(H(\gamma,R))=\sigma(R)$ for all preference profiles $R\in\mathcal{R}$ and some \hyperref[l]{choice liberal} rule $\sigma\in\Sigma_l$ such that for all players $i\in N$ and all numbers $x,y\in\{2i,2i-1\}$ such that $x\neq y$, $\omega_x P_i\omega_y$ implies $\omega_x\in\sigma(R)$ and $\omega_y\notin\sigma(R)$.
    
    The first step is to show by contradiction that for any outcome $\omega\in\{\omega_1,\dots,\omega_4\}$, there exists some player $i\in\{1,2\}$ such that $\omega\in\eta(S_i,s_j)$ for every opponent's strategy $s_j\in S_j$. Suppose not. Then, there exists some strategy profile $s\in S$ and some outcome $\omega\in\{\omega_1,\dots,\omega_4\}$ such that $\omega\notin\eta(S_i,s_j)$ for all players $i\in\{1,2\}$. Fix that outcome $\omega\in\{\omega_1,\dots,\omega_4\}$ and that strategy profile $s\in S$. Then, let the preference profile $R\in\mathcal{R}$ satisfy $\omega P_i\eta(s)R_i\omega'$ for all players $i\in\{1,2\}$ and all outcomes $\omega'\notin\{\omega,\eta(s)\}$, where $\omega=\omega_{2i}$ and $\eta(s)=\omega_{2i-1}$ for some player $i\in\{1,2\}$. Then, $s\in H(\gamma,R)$ yet $\eta(s)\notin\sigma(R)$, thus contradicting the fact that $\eta(H(\gamma,R))=\sigma(R)$ for all preference profiles $R\in\mathcal{R}$.
    
    The second step is to show by contradiction that for all outcomes $\omega,\omega'\in\{\omega_1,\dots,\omega_4\}$, if $\omega\in\eta(S_i,s_j)$ for all opponent's strategies $s_j\in S_j$, and $\omega'\in\eta(S_j,s_i)$ for all opponent's strategies $s_i\in S_i$; then $i=j$. Given any outcome $\omega\in\{\omega_1,\dots,\omega_4\}$, let $i(\omega)\in \{1,2\}$ be the player $i\in \{1,2\}$ for whom $\omega\in\eta(S_i,s_j)$ for all opponent's strategies $s_j\in S_j$. Suppose that for some pair of outcomes $\omega,\omega'\in\{\omega_1,\dots,\omega_4\}$, we have $i(\omega)\neq i(\omega')$. Then, $H(\gamma,R)=\emptyset$ whenever the preference profile $R\in\mathcal{R}$ satisfies $\omega P_{i(\omega)}\omega''$ for all outcomes $\omega''\neq\omega$ as well as $\omega'P_{i(\omega')}\omega''$ for all outcomes $\omega''\neq\omega'$, thus contradicting the fact that $H(\gamma,R)\neq\emptyset$ for all preference profiles $R\in\mathcal{R}$.

    The third step is to show by contradiction that $\Sigma_n\cap\Sigma_l=\emptyset$. Consider the unique player $i\in\{1,2\}$ satisfying $i(\omega)$ for all outcomes $\omega\in\{\omega_1,\dots,\omega_4\}$. Then, let the preference profile $R\in\mathcal{R}$ satisfy $\omega_{2j} P_i\omega$ for all outcomes $\omega\neq\omega_{2j}$, and $\omega_{2j-1}P_j\omega_{2j}P_j\omega$ for all outcomes $\omega\notin\{\omega_{2j-1},\omega_{2j}\}$. Fix any strategy $s_i\in S_i$ satisfying $\omega_{2j-1}\notin\eta(S_j,s_i)$. Then, $s=(s_i,s_j)\in H(\gamma,R)$ implies $\eta(s)=\omega_{2j}$. Thus, $\omega_{2j}\in \eta(H(\gamma,R))=\sigma(R)$ yet $\omega_{2j}\notin\sigma(R)$, which is a contradiction. Therefore, $\Sigma_n\cap\Sigma_l=\emptyset$.

\begin{statement}
    If $n\geqslant 3$ and $|\Omega|\geqslant 2n$, then $\Sigma_l\cap\Sigma_n\neq\emptyset$.
\end{statement}

    Let $n\geqslant 3$ and $|\Omega|\geqslant 2n$.

\begin{claim}
    $(\Sigma_l\neq\emptyset)$.
\end{claim}

    Let $\Omega=\{\omega_1,\dots,\omega_m\}$, and let the social choice rule $\lambda:\mathcal{R}\to2^\Omega\backslash\{\emptyset\}$ satisfy, for all preference profiles $R\in\mathcal{R}$, 
\begin{gather}\label{key0}
    \lambda(R)=\big\{\omega_x\in\Omega\mid(\exists i\in N)[(\forall x,y\in\{2i,2i-1\})(x\neq y)](\omega_xR_i\omega_y)\big\}
\end{gather}
    Thus, the social choice rule $\lambda:\mathcal{R}\to2^\Omega\backslash\{\emptyset\}$ associates a unique pair of outcomes with every player $i\in N$, and adds to its image exactly the one preferred by every player $i\in N$. To see that $\lambda\in\Sigma_l$, pick any pair $(i,R)\in N\times\mathcal{R}$. Let $x,y\in\{2i,2i-1\}$, where $x\neq y$. If $\omega_x P_i\omega_y$, then $\omega_xR_i\omega_y$ and $\omega_y\neg R_i\omega_x$. Thus, by \cref{key0}, $\omega_x\in\lambda(R)$ and $\omega_y\notin\lambda(R)$. Hence, $\lambda\in\Sigma_l$.

\begin{claim}
    $(\lambda\in\Sigma_n)$.
\end{claim}
    
    The proof is constructive. Let the mechanism $\gamma=(S,\eta)\in\Gamma$ satisfy, for all players $i\in N$,
\begin{gather}
    s_i=(z_i,a_i)\in S_i=\{\omega_{2i},\omega_{2i-1}\}\times\{1,\dots,n\}
\end{gather}    
    Further, let the outcome function $\eta:S\to\Omega$ satisfy, for all strategy profiles $s\in S$,
\begin{gather}
\eta(s)=
\begin{cases}
    z_k & \text{if }[(\forall i\in N\backslash\{j\})(a_i=k)]\vee[(\forall i\in N)(a_i=k)]\\
    \min\left\{z_i\in\Omega\mid i\in N\right\} &\text{else}
\end{cases}
\end{gather}
    In this mechanism, all players simultaneously name one player and one outcome from the pair they are decisive over. When either all or all but one players name the same player, the outcome named by the commonly chosen player is the final outcome; else, the final outcome is the one named by the player with the lowest index.

\begin{lemma}
    $(\forall R\in\mathcal{R})(\eta(H(\gamma,R))\subseteq \lambda(R))$.
\end{lemma}

    The proof is by contraposition. Consider any preference profile $R\in\mathcal{R}$ and any strategy profile $s\in S$ satisfying $\eta(s)\notin\lambda(R)$. Then, there exists some player $i\in N$ such that $\omega_x P_i\omega_y$, where $x,y\in\{2i,2i-1\}$, $x\neq y$ and $\omega_y=\eta(s)$. Fix that player $i\in N$. Then, $s_i=(\omega_y,a_i)$. Let the strategy $s_i'\in S_i$ satisfy $s_i'=(\omega_x,a_i)$. Then, $\eta(s_i',s_{-i})=\omega_x$. But since $\omega_xP_i\omega_y$, it follows that $\eta(s'_i,s_{-i})P_i\eta(s)$. Thus, $s\notin H(\gamma,R)$. Therefore, $\eta(H(\gamma,R))\subseteq \lambda(R)$ for all preference profiles $R\in\mathcal{R}$.

\begin{lemma}
    $(\forall R\in\mathcal{R})(\lambda(R)\subseteq\eta(H(\gamma,R)))$.
\end{lemma}
    
    The proof is direct. Consider any preference profile $R\in\mathcal{R}$ and any outcome $\omega\in\lambda(R)$. Fix the player $j\in N$ for whom $\omega\in\{\omega_{2j},\omega_{2j-1}\}$. Then, let the strategy profile $s\in S$ satisfy $s_j=(\omega,j)$ and $s_i=(z_i,j)$ for all players $i\neq j$. Then, $\eta(s)=\omega$. Since $\omega\in\lambda(R)$, it follows that $\omega R_j\omega'$, where $\omega'\in\{\omega_{2j},\omega_{2j-1}\}\backslash\{\omega\}$. Let the strategy $s_j'=(z_j',a_j')\in S_j$ satisfy $z_j'\neq\omega$ or $a_j'\neq a_j$. In the former case, $\eta(s_j',s_{-j})=z_j'$; in the latter case, $\eta(s_j',s_{-j})\in\{\omega,z_j'\}$. Since $\omega R_jz_j'$, it follows that $\eta(s)R_j\eta(s_j',s_{-j})$ for all strategies $s_j'\in S_j$. Moreover, $\eta(s_i',s_{-i})=\eta(s)$ for all strategies $s_i'\in S_i$ and all players $i\neq j$. Thus, $s\in H(\gamma,R)$. Therefore, $\lambda(R)\subseteq\eta(H(\gamma,R))$ for all preference profiles $R\in\mathcal{R}$.
\end{proof}

\section{Examples}\label{sec.ex}

\Cref{sec.ex} provides some examples inspired by \citeauthor{sen_70}'s (\citeyear{sen_70}, p. 155) that illuminate the three exhaustive cases considered in the proof of \Cref{th.cl}: \Cref{subsec.ex1} considers the case in which there are at least three players but not enough outcomes, \Cref{subsec.ex2} considers the case in which there are three players and enough outcomes, and \Cref{subsec.ex3} considers the two-player case.

\subsection{Three Players \& Four Outcomes}\label{subsec.ex1}

Consider a group of three classmates: \emph{Alice} ($A$), \emph{Bob} ($B$) and \emph{Carl} ($C$); and Mas-Colell's et al. ``\emph{Microeconomic Theory}'' textbook. Their problem is deciding who reads the book to prepare for their upcoming exam. Then, either \emph{only $A$ reads it} ($\omega_1$), \emph{only $B$ reads it} ($\omega_2$), \emph{only $C$ reads it} ($\omega_3$), or \emph{nobody reads it} ($\omega_4$). Consider a social choice rule $\sigma:\mathcal{R}\to2^\Omega\backslash\{\emptyset\}$ that lets each of them decide whether to read the book when nobody else is reading it: namely, it lets $A$ be decisive over the pair $\{\omega_1,\omega_4\}$, lets $B$ be decisive over the pair $\{\omega_2,\omega_4\}$, and lets $C$ be decisive over the pair $\{\omega_3,\omega_4\}$. Consider a preference profile $R\in\mathcal{R}$ in which both $A$ and $B$ want to read the book but $C$ does not; so that $\omega_1P_A\omega_4$, $\omega_2P_B\omega_4$ and $\omega_4P_C\omega_3$. Then, by \hyperref[l]{choice liberalism}, $\omega_4\in\sigma(R)$ and $\omega_4\notin\sigma(R)$, which is a contradiction. 

\subsection{Three Players \& Six Outcomes}\label{subsec.ex2}

Suppose now there are two copies of the book, so that either \emph{only $A$ reads it} ($\omega_1$), \emph{only $B$ reads it} ($\omega_2$), \emph{only $C$ reads it} ($\omega_3$), \emph{only $A$ and $B$ read it} ($\omega_4$), \emph{only $A$ and $C$ read it} ($\omega_5$), or \emph{only $B$ and $C$ read it} ($\omega_6$). Consider a social choice rule $\sigma:\mathcal{R}\to2^\Omega\backslash\{\emptyset\}$ that lets $A$ decide whether to read the book when $B$ does and $C$ does not, lets $B$ decide whether to read the book when $C$ does and $A$ does not, and lets $C$ decide whether to read the book when $A$ does and $B$ does not. Hence, $A$ is decisive over the pair $\{\omega_2,\omega_4\}$, $B$ is decisive over the pair $\{\omega_3,\omega_6\}$, and $C$ is decisive over the pair $\{\omega_1,\omega_5\}$. Further, consider a preference profile $R\in\mathcal{R}$ in which all of them want to read the book, so that $\omega_4 P_A\omega_2$, $\omega_6 P_B\omega_3$ and $\omega_5 P_C\omega_1$. Then, by \hyperref[l]{choice liberalism}, $\sigma(R)=\{\omega_4,\omega_5,\omega_6\}$.

Now, consider this adaptation of the mechanism provided in the proof of \Cref{th.cl}:
\begin{itemize}
    \item $A$ simultaneously names either $\omega_2$ or $\omega_4$ and either $A$, $B$ or $C$, 
    \item $B$ simultaneously names either $\omega_3$ or $\omega_6$ and either $A$, $B$ or $C$,
    \item $C$ simultaneously names either $\omega_1$ or $\omega_5$ and either $A$, $B$ or $C$.
\end{itemize}
If \emph{at least} all but one classmates name the same classmate, the final outcome is the one named by that classmate; otherwise, the final outcome is---without loss of generality---the one named by $A$.

Consider any strategy profile in which $A$ names $\omega_4$, and $A$, $B$ and $C$ all name $A$. Then, the final outcome is the one named by $A$; namely, $\omega_4$. If $A$ changes her strategy and names $\omega_2$ instead, the final outcome is $\omega_2$; but since $\omega_4P_A\omega_2$, $A$ cannot profitably deviate by doing so. Suppose that some classmate names someone else but $A$. Then, the final outcome remains $\omega_4$, so no one can profitably deviate by doing so. Hence, $\omega_4$ is a Nash equilibrium outcome. A similar reasoning applies to the other socially desirable outcomes $\omega_5,\omega_6$. Now, consider any outcome $\omega\in\{\omega_1,\omega_2,\omega_3\}$. If $\omega_1$ is the final outcome, then $C$ is naming $\omega_1$. But if he named $\omega_5$ instead, the final outcome would be $\omega_5$. And since $\omega_5P_C\omega_1$, naming $\omega_5$ is a profitable deviation. A similar reasoning applies to the other socially non-desirable outcomes $\omega_2,\omega_3$. 

\subsection{Two Players}\label{subsec.ex3}

Suppose there are only Ann ($A$) and Bob ($B$) and two textbooks: Mas-Colell's et al. ``\emph{Microeconomic Theory}'' and Varian's ``\emph{Intermediate Microeconomics}''. Then, they need to decide whether \emph{$A$ reads Mas-Colell and $B$ reads Varian} ($\omega_1$), \emph{$A$ reads Varian and $B$ reads Mas-Colell} ($\omega_2$), \emph{$A$ reads Mas-Colell and $B$ reads nothing} ($\omega_3$), \emph{$A$ reads nothing and $B$ reads Mas-Colell} ($\omega_4$), \emph{$A$ reads Varian and $B$ reads nothing} ($\omega_5$), or \emph{$A$ reads nothing and $B$ reads Varian} ($\omega_6$). Suppose that $\sigma:\mathcal{R}\to2^\Omega\backslash\{\emptyset\}$ is a social choice rule that lets both $A$ and $B$ decide whether to read Mas-Colell or nothing when the other one reads Varian. Then, $A$ is decisive over the pair $\{\omega_1,\omega_6\}$, and $B$ is decisive over the pair $\{\omega_2,\omega_5\}$. Suppose further that both of them prefer to read Mas-Colell over reading nothing, so that $\omega_1 P_A\omega_6$ and $\omega_2 P_B\omega_5$. Then, by \hyperref[l]{choice liberalism}, $\{\omega_1,\omega_2\}\subseteq\sigma(R)$ and $\{\omega_5,\omega_6\}\cap\sigma(R)=\emptyset$. Suppose, without loss of generality, that $\sigma(R)=\{\omega_1,\omega_2\}$.

Since every two-player mechanism---no matter how complex---can be represented by a matrix; let $A$ be the \emph{row player}, and let $B$ be the \emph{column player}. Any mechanism that \hyperref[ni]{Nash implements} $\sigma:\mathcal{R}\to 2^{\Omega}\backslash\{\emptyset\}$ must have a Nash equilibrium $s=(s_A,s_B)$ leading to outcome $\omega_1$. Then, $B$ cannot have any strategy $s'_B\neq s_B$ such that $(s_A,s_B')$ yields some outcome $\omega'\neq\omega_1$; for if $\omega_1$ were $B$'s most disliked outcome, he would have a profitable deviation by choosing $s_B'$. Then, the matrix associated with the mechanism must have some row containing only $\omega_1$. 

Moreover, any mechanism that \hyperref[ni]{Nash implements} $\sigma:\mathcal{R}\to 2^{\Omega}\backslash\{\emptyset\}$ must also have a Nash equilibrium $s'=(s'_A,s'_B)$ leading to outcome $\omega_2$. Consider any such strategy profile $s'=(s_A',s_B')$. Since the matrix has a row containing only $\omega_1$, $A$ must have some strategy $s_A''\neq s'_A$ such that $(s_A'',s_B')$ leads to outcome $\omega_1$. If it were the case that $\omega_1P_A\omega_2$, then $A$ would have a profitable deviation by choosing strategy $s_A''$. But then, no strategy profile yielding $\omega_2$ could be a Nash equilibrium. And hence, the social choice rule $\sigma:\mathcal{R}\to 2^{\Omega}\backslash\{\emptyset\}$ is not \hyperref[ni]{Nash implementable}. 

\acknowledgments{I am sincerely thankful to Fatma Aslan, Vicen\c c Esteve Guasch, Ben McQuillin, R\'{o}bert Somogyi, Christopher Stapenhurst, Mich Tvede and four anonymous referees for providing several comments that substantially improved this note. All errors are only mine. No generative AI was used in the write-up of this paper. I acknowledge no conflict of interest.}

\conflictofinterest{I declare no conflict of interest.}

\dataavailability{No data was used to write this paper.}

\funding{None.}

\printbibliography[]
\end{document}